\documentclass[doublecol]{epl2} 

\title{Dynamic susceptibility and dynamic correlations in spin ice }

\author{M. I. Ryzhkin\inst{1} \and I. A. Ryzhkin\inst{1} \and S. T. Bramwell\inst{2}}
\shortauthor{M. I. Ryzhkin \etal}

\institute{                    
  \inst{1}Institute of Solid State Physics RAS - 2 Academician Ossipyan street, Chernogolovka, Moscow region 142432, Russia\\
  \inst{2}London Centre for Nanotechnology and Department of Physics and Astronomy, University College of London - 17-19  Gordon Street, London WC1H0AJ, UK}


\abstract{
Here we calculate the dynamic susceptibility and dynamic correlation function in spin ice using the model of emergent magnetic monopoles. Calculations are based on a method originally suggested for the description of dynamic processes in water ice (non-equilibrium thermodynamics approach). We show that for $T\rightarrow 0$  the dynamic correlation function reproduces the transverse dipole correlations (static correlation function) characteristic of spin ice in its ground state. At non-zero temperatures the dynamic correlation function includes an additional longitudinal component which decreases as the temperature decreases. Both terms (transverse and longitudinal) exhibit identical Debye-like dependences on frequency but with different relaxation times: the magnetic Coulomb interaction of monopoles reduces the longitudinal relaxation time with respect to the transverse one.  We calculate the dielectric function for the magnetic monopole gas and discuss how the non-equilibrium thermodynamics approach exposes corrections to the  Debye-H\"uckel theory of magnetic monopoles and the concept of ``entropic charge''. }

\begin{document}

\maketitle

\section{Intoduction}

Spin ice is the name given to compounds such as $\rm Ho_2Ti_2O_7, Dy_2Ti_2O_7$ which demonstrate unusual magnetic correlations~\cite{b.1}.  The magnetic ions ${\rm Ho^{3+}}$ and ${\rm Dy^{3+}}$ sit at the vertices of regular tetrahedra linked into a three-dimensional pyrochlore lattice. Due to strong anisotropy the atomic magnetic moments (spins) of the magnetic ions can be directed only along local anisotropic axes connecting the centers of nearest tetrahedra. The ground state is characterized by the ice rule: two spins of each tetrahedron are directed toward its center, and two other spins, away from its center (see fig.1). This rule leads to a degeneracy that diverges exponentially with the number of spins ~\cite{b.2,b.3} and to implicit topological ordering. To illustrate this it is useful to  present a ground state configuration as a set of strongly entangled strings (one of them is shown in fig.1). Each such string is a line drawn through lattice bonds with magnetic ions along spins, and the ice rule ensures that strings can either be closed or they end at the sample boundary. Spins are strongly correlated along strings, therefore long strings contributed to long-range correlations, short and closed strings destroy these correlations. The final result depends on the relative proportion of long and short strings. From the string picture it becomes obvious that there might be  long-range correlations between spins in spin ice.\\
\indent Those correlations between spins can be characterized by an equilibrium (or static) correlation function for the local magnetization. As was shown in~\cite{b.4,b.5,b.6} by averaging over all ground state configurations this function has a striking dipole like form 

\begin{equation}
\label{eq.1}
S_{\alpha\beta}(\vect{q})=\langle M_\alpha (\vect{q})M_\beta \vect{-q})\rangle \propto \Bigl( \delta_{\alpha\beta}-\frac{q_\alpha q_\beta}{q^2}\Bigr)
\end{equation}

From the same string picture it is obvious that any ground state configuration is frozen: one cannot change it without the breaking some of strings or without violating the ice rule. The flipping of a spin in the ground state breaks the ice rule in two neighboring tetrahedra: one tetrahedron has spins directed three-in and one-out, while its neighbor has one-in and three-out. These tetrahedra can be considered as positive and negative emergent magnetic monopoles respectively~\cite{b.7,b.8}.  By means of further spin flips the magnetic monopoles can move through the lattice and thereby change the spin configuration (see fig.2). The magnetic monopole model affords an economical way of describing the dynamic processes in spin ice, as one substitutes the strongly correlated system of spins by one of dilute and weakly interacting magnetic monopoles.

\begin{figure}
\includegraphics[height=62mm]{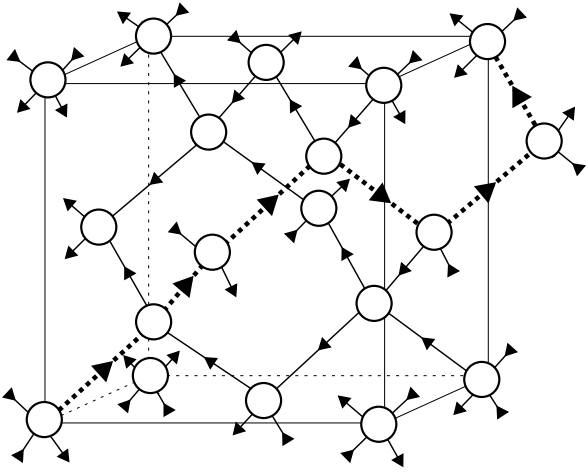}
\caption{Local spin ordering in spin ice: spins are directed along local anisotropy axes according the ice rule (two spins in and two spins out per vertex). The dotted line is an example of a long string that contributes to long-range correlations.}
\label{fig.1}
\end{figure}

\indent Here we use this model to calculate the dynamic susceptibility and dynamic correlation function of spin ice at non-zero temperatures. We also discuss their relationship with static analogs obtained for the ground state and with results of other works.
\vspace{1mm}
\section{Basic equations}

To calculate the dynamic susceptibility we find the magnetization of spin ice in a time dependent and inhomogeneous magnetic field using Jaccard's method originally suggested for electric relaxation in common water ice~\cite{b.9}. Its detailed description, corresponding to the notations of the present paper, can be found in~\cite{b.10}. In this method fluxes and concentrations of magnetic monopoles, magnetization and magnetic field are defined by the following set of equations:

\begin{eqnarray}
\label{eq.2}
\vect{j_\pm}+D_\pm\nabla\delta n_\pm &=&  \frac{\sigma_\pm}{Q_\pm^2}\Bigl[Q_\pm(\vect{H}+\vect{h})-\eta_\pm\Phi\vect{\Omega}\Bigr]\\
\partial\vect{\Omega}/\partial t &=& \sum_{\pm}\eta_\pm \vect{j_\pm}\\
\partial\delta n_\pm/\partial t &=&-\nabla\cdot\vect{j_\pm}\\
\nabla\cdot \vect{h}&=&4\pi\sum_{\pm}Q_\pm\delta n_\pm
\end{eqnarray}

\begin{figure}
\includegraphics[height=62mm]{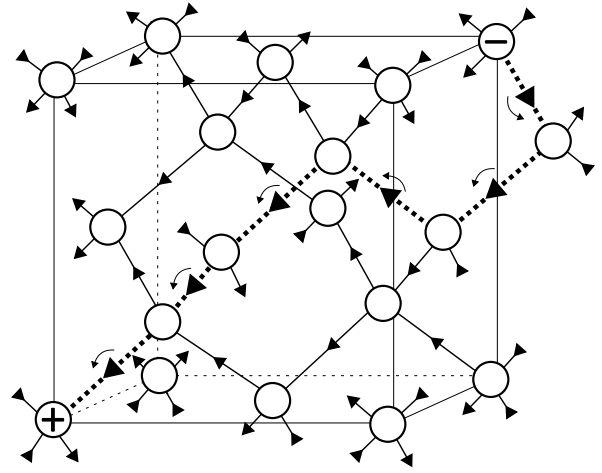}
\caption{Flips of spins along the string shown in fig.1 create a pair of magnetic monopoles bounded by the string (monopoles shown as plus and minus). The movement of a monopole along the string changes the spin configuration on it.}
\label{fig.2}
\end{figure}

Equations (2) define linear response fluxes of magnetic monopoles due to applied disturbances. Here $\sigma_\pm,D_\pm, Q_\pm=\pm Q$ are specific conductivities, diffusion coefficients, and magnetic charges of positive and negative magnetic monopoles respectively, $\eta_\pm=\pm 1$ describe signs and type of magnetic ordering due to monopole fluxes. Fluxes are induced by the external magnetic field $\vect{H}$ and by the magnetic field generated by magnetic monopoles $\vect{h}$. The configuration vector   $\vect{\Omega}$ and concentration gradients also contribute to the fluxes. A physical sense of the configuration vector is very simple, it is proportional to the magnetization $\vect{M}=Q\vect{\Omega}$. Equations (2) are close analogs to ones for fluxes of electrons and holes in semiconductors with the exception of the entropic term proportional to $\vect{\Omega}$. A parameter $\Phi =8ak_BT/\sqrt{3}$  , calculated in~\cite{b.11}, can be considered as an entropic charge interaction~\cite{b.12}.

\indent Equation (3) defines a modification of the spin ordering by magnetic monopole fluxes (see the spins along dotted lines in fig. 1,2), eqs. (4) are continuity equations, and eq. (5) is the magnetic analog of Poisson's equation. Specific conductivities and diffusion coefficients are related by the expressions $\sigma_\pm=Q_\pm^2D_\pm n_{0\pm}/k_B T=Q\mu_\pm n_{0\pm}$, where $n_{0\pm}=n_0$  and $\mu_\pm$ are the concentrations and mobilities of monopoles respectively. Note that eq. (4) is approximate in that it excludes chemical kinetic terms that describe the creation and annihilation of magnetic monopoles, as well as bound pair formation.\\
\indent In the Fourier presentation eqs.(2-5) become linear algebraic ones, and they can be written in the forms (we use the same notation for variables in Fourier presentation):
\begin{eqnarray}
\vect{j_\pm}+D_\pm i\vect{q}\delta n_\pm &=& \pm \frac{\sigma_\pm}{Q^2}\Bigl[Q(\vect{H}+\vect{h})-\Phi\vect{\Omega}\Bigr]\\
-i\omega\vect{\Omega} &=&\vect{j_+}-\vect{j_-}\\
\omega\delta n_\pm &=&\vect{q}\cdot\vect{j_\pm}\\
i\vect{q}\cdot \vect{h}&=&4\pi Q(\delta n_+ - \delta n_-)
\end{eqnarray}

\section{Results}
Eliminating first $\delta n_\pm$, then $\vect{j_\pm}$, we find the components of magnetization $M_\alpha$ and magnetic field $h_\alpha$.  Skipping simple calculations we give final expressions for the case of equal diffusion coefficients $D=D_+=D_-$:

\begin{eqnarray}
M_\alpha&=&\frac{Q^2/(\Phi\tau)}{1/\tau-i\omega}\Bigl(\delta_{\alpha\beta} -\frac{q_\alpha q_\beta}{q^2}\Bigr)+\nonumber\\
             &   &\frac{Q^2/(\Phi\tau)}{1/\tau'-i\omega+Dq^2}\frac{q_\alpha q_\beta}{q^2 }H_\beta
\end{eqnarray}

\begin{equation}
h_\alpha=-\frac{4\pi Q^2/(\Phi\tau)}{1/\tau'-i\omega+Dq^2}\frac{q_\alpha q_\beta}{q^2 }H_\beta
\end{equation}\\
where transverse and longitudinal relaxation times are defined by the following equations respectively:

\begin{equation}
\frac{1}{\tau}=\frac{2\Phi Dn_0}{k_BT}, \quad     \frac{1}{\tau'}=\frac{2(4\pi Q^2+\Phi) Dn_0}{k_BT} 
\end{equation}\\
From the eq.(10) we get the following expression for a commonly used magnetic susceptibility:

\begin{eqnarray}
\chi_{\alpha\beta} (\vect{q},\omega)&=&\frac{Q^2/(\Phi\tau)}{1/\tau-i\omega}\Bigl(\delta_{\alpha\beta} -\frac{q_\alpha q_\beta}{q^2}\Bigr) \nonumber\\
                                                     &  &+\frac{Q^2/(\Phi\tau)}{1/\tau'-i\omega+Dq^2}\frac{q_\alpha q_\beta}{q^2 }
\end{eqnarray}

This susceptibility relates external field and generated magnetizations, but there are well-founded reasons that real experiments - in particular neutron scattering - must be described by the generalised susceptibility which gives the response to the bulk field $\tilde{H}_\beta$:\\
\begin{equation}
M_\alpha=\tilde{\chi}_{\alpha\beta} (\vect{q},\omega)\tilde{H}_\beta=\tilde{\chi}_{\alpha\beta} (\vect{q},\omega)(H_\beta+h_\beta)
\end{equation}\\
Using eqs.(10, 11) it easy to get for $\tilde{\chi}_{\alpha\beta} (\vect{q},\omega)$ the following expression:
\begin{eqnarray}
\tilde{\chi}_{\alpha\beta} (\vect{q},\omega)&=&\frac{Q^2/\Phi}{1-i\omega\tau}\Bigl(\delta_{\alpha\beta} -\frac{q_\alpha q_\beta}{q^2}\Bigr) \nonumber\\
                                                     &  &+\frac{Q^2/\Phi}{1-i\omega\tau+\tau Dq^2}\frac{q_\alpha q_\beta}{q^2 }
\end{eqnarray}\\
Note that the generalised susceptibility (15) can be found from the common susceptibility by equating the relaxation times $\tau'\rightarrow\tau$. In the following we consider the common susceptibility $\chi_{\alpha\beta}$, and respective results for the generalised one can be got by means of simple substitution  $\tau'\rightarrow\tau$.\\
\indent Using eq.(13) and the fluctuation-dissipation theorem in its classical form
\begin{equation}
{\rm Im}\chi_{\alpha\beta}(\vect{q},\omega)=\frac{\omega}{2k_BT}S_{\alpha\beta}(\vect{q},\omega)
\end{equation}

\noindent we get the dynamical correlation function $S_{\alpha\beta}(\vect{q},\omega)$. Omitting simple computations we get:
\begin{eqnarray}
S_{\alpha\beta} (\vect{q},\omega)&=&4Q^2Dn_0\Bigl[\frac{1}{1/\tau^2+\omega^2}\Bigl(\delta_{\alpha\beta}-\frac{q_\alpha q_\beta}{q^2}\Bigr)+ \nonumber\\
                                              &   &\frac{1}{(1/\tau'+Dq^2)^2 +\omega^2}\frac{q_\alpha q_\beta}{q^2}\Bigr]
\end{eqnarray}

The respective formula from the generalised correlation function has the form
\begin{eqnarray}
\tilde{S}_{\alpha\beta} (\vect{q},\omega)&=&4Q^2Dn_0\Bigl[\frac{1}{1/\tau^2+\omega^2}\Bigl(\delta_{\alpha\beta}-\frac{q_\alpha q_\beta}{q^2}\Bigr)+ \nonumber\\
                                              &   &\frac{1}{(1/\tau+Dq^2)^2 +\omega^2}\frac{q_\alpha q_\beta}{q^2}\Bigr]
\end{eqnarray}

\section{Discussion}
Now let us discuss the physical sense of the obtained results, compare them with results of other papers and with experiments.\\
\indent First, we discuss the fundamental question: does the model of emergent monopoles agree with Maxwell’s equations?  The answer is “yes”. Indeed, the applied field is solenoidal $q_\alpha H_\alpha=0$. Then we have to write an induction of magnetic field in the form:
\begin{equation}
B_\alpha=H_\alpha+4\pi M_\alpha + h_\alpha=H_\alpha+\frac{Q^2/\Phi}{1-i\omega\tau}\Bigl(\delta_{\alpha\beta} -\frac{q_\alpha q_\beta}{q^2}\Bigr)
\end{equation}
Therefore the magnetic induction is also solenoidal, $q_\alpha B_\alpha=0$, in agreement with Maxwell’s equation $\nabla\cdot\vect{B}=0$ (there are no genuine magnetic monopoles). The condition is satisfied due to the exact cancellation of longitudinal terms from eqs.(10, 11) in eq.(19). Therefore the accounting of the field $\vect{h}$ into eq. (2) is fundamentally important: without this field one comes to misleading conclusion that genuine magnetic monopoles exist. The exact cancellation of longitudinal terms from $M_\alpha$ and  $h_\alpha$ is always maintained: for different diffusion coefficients and at all values of temperature. Also note that the term in eq. (19), after the first equation sign, may seem unusual in comparison with common definitions of magnetic induction. Why must we include the supplementary term $h_\alpha$? In fact this term arises from nearly free monopoles, and in some sense it is similar to the contribution of free electrons to an electrical induction in semiconductors.\\ 
\indent Second, we note that the magnetization in (10) includes both transverse and longitudinal terms.  Both terms have Debye dependence on frequency, but with different relaxation times. The first (transverse) term takes a transverse relaxation time $\tau$, the second (longitudinal) one takes a longitudinal time $\tau'$, see eqs.(12). A ratio of times can be estimated as $\tau/\tau'\approx 1+8.3/T \gg 1$  for low temperature $T<2K$ In fact we can assert that the magnetic monopole model works well only in this low temperature region, where the concentration of monopoles is low (much lower than the concentration of regular tetrahedrons without violations of ice rules). At higher temperatures the concentration of magnetic monopoles is too high to consider them as weakly interacting (independent) quasiparticles. Note that we take into account the Coulomb interaction between monopoles via the field $\vect{h}$, and that is a kind of mean field approximation. But almost all of the starting equations hold only for the low concentration limit. We stress: this flaw is not specific feature of our approach, but strictly it is inherent to any theory which uses a magnetic monopole picture.\\
\indent Third, for time dependent correlation functions we get the following expressions:
\begin{eqnarray}
S_{\alpha\beta} (\vect{q},t)&=&\frac{\sqrt{3}Q^2}{8a}\Bigl[\Bigl(\delta_{\alpha\beta}-\frac{q_\alpha q_\beta}{q^2}\Bigr)e^{-t/\tau}+ \nonumber\\
                                              &   &\frac{e^{-(1/\tau'+Dq^2)t}}{(\tau/\tau'+\tau Dq^2)^2 +\omega^2}\frac{q_\alpha q_\beta}{q^2}\Bigr]
\end{eqnarray}
\begin{eqnarray}
\tilde{S}_{\alpha\beta} (\vect{q},t)&=&\frac{\sqrt{3}Q^2}{8a}\Bigl[\Bigl(\delta_{\alpha\beta}-\frac{q_\alpha q_\beta}{q^2}\Bigr)e^{-t/\tau}+ \nonumber\\
                                              &   &\frac{e^{-(1/\tau+Dq^2)t}}{(1+\tau Dq^2)^2 +\omega^2}\frac{q_\alpha q_\beta}{q^2}\Bigr]
\end{eqnarray}\\
In the limit $T\rightarrow 0$   for all $t>0$ we come to the equation
\begin{equation}
S_{\alpha\beta} (\vect{q},t)=\tilde{S}_{\alpha\beta} (\vect{q},t)=\frac{\sqrt{3}Q^2}{8a}\Bigl[\Bigl(\delta_{\alpha\beta}-\frac{q_\alpha q_\beta}{q^2}\Bigr)
\end{equation}
which coincides with the static correlation function~\cite{b.4,b.5,b.6}. For coincidence with~\cite{b.4} one should take parameter $\lambda=1$. The independence on time of (22) means that at zero temperature there are no relaxation processes. But if we first tend $t\rightarrow\infty$  we come to the different result:
\begin{equation}
S_{\alpha\beta} (\vect{q},t)=\tilde{S}_{\alpha\beta} (\vect{q},t)=0
\end{equation}
\noindent That is if there are relaxation processes (correlation functions tend to zero as time tends to infinity).\\
\indent Finally we compare the formulae (20, 21) with results from the works~\cite{b.13,b.14}. The eq. (21) looks exactly like the result of~\cite{b.13,b.14} for Heisenberg's model if we equalize:
\begin{equation}
8\Gamma \lambda T\rightarrow \tau^{-1}, \quad  8\Gamma J a^2 \rightarrow D, \quad \xi^2 \rightarrow D\tau
\end{equation}
But the eq. (20) has a slightly different form. There are two characteristic lengths in (20): the diffusion length $\xi_{diff}=\xi=(\tau D)^{1/2}$ and the Debye screening length $\xi_D=\xi/(4\pi Q^2/\Phi +1)^{1/2}$ . This difference is a consequence of neglecting the Coulomb interaction between monopoles (that is by neglecting the field $h_\alpha$) in~\cite{b.13,b.14}. In fact a magnetic Coulomb interaction leads to the inequality $\xi_{diff} \gg \xi_D$ for $T<2K$.\\ 
\indent The longitudinal susceptibility was considered by one of us in a recent article~\cite{b.15}. Our present result for the longitudinal susceptibility coincides with that of~\cite{b.15} one in the limit  and differs from it in general case. But for generalised susceptibility the both results exactly coincide.\\
\indent To explore the relationship with Debye-H\"uckel theory~\cite{b.12} it is interesting to calculate an effective dielectric function. We start by defining a longitudinal common susceptibility from eq.(14) for $\omega=0$:
\begin{equation}
\chi^L(q)=\frac{Q^2/\Phi}{1+4\pi Q^2/\Phi+\tau Dq^2}=\frac{\chi_T \xi^{-2}}{\xi^{-2}_D +q^2}
\end{equation}
where $\chi_T\equiv \tilde{\chi}(q=0,\omega=0)$   is the isothermal bulk susceptibility. Next we consider the response of a system to an applied longitudinal field, as might arise in practice from a fixed system of effective magnetic charges. The longitudinal static dielectric function may be defined by:
\begin{equation}
\frac{1}{\epsilon^L(q)}=\frac{H(q)+h(q)}{H(q)}=\frac{H(q)-4\pi M(q)}{H(q)}=\frac{\xi^{-2}+q^2}{\xi^{-2}_D +q^2}
\end{equation}\\
The screened Coulomb potential per unit charge is given by:
\begin{equation}
\phi(r)=FT\Biggl[\frac{1}{q^2 \epsilon^L (q)}\Biggr]=\frac{2}{\pi}\int_{0}^{\infty}\frac{\sin(qr)}{qr\epsilon^L(q)}dq
\end{equation}
\noindent The result is: 
\begin{equation}
\phi(r)=\frac{e^{-r/\xi_D}+1/4\pi\chi_T}{r(1+1/4\pi\chi_T)}
\end{equation}
When $\xi_D$ is large, this becomes the Coulomb form,$1/r$. When $\chi_T$  is large, it becomes the Debye-H\"uckel form, but at long distances there is a crossover to a modified Coulomb form  $A/r$ with reduced amplitude $A=1/(1+4\pi\chi_T)$. The monopoles cannot completely screen each other ~\cite{b.16}. It is easily shown that the positive charge induced by a negative charge at the origin is less than a whole charge $Q$ by a factor $1/(1+ 1/4 \pi \chi_T)$. Note that eq. (28) applies for large $\xi_D$   only: for small $\xi_D$  the cutoff given by the lattice constant must be accounted for, but this is beyond the scope of the theory discussed here. In analysis of the experimental specific heat of $\rm Dy_2Ti_2O_7$~\cite{b.12} the cutoff dominates at temperatures above $\approx 1K$, where present theory also becomes inaccurate.\\
\indent It is also interesting to discuss the relation of the present theory to the concept of entropic charge~\cite{b.12}. If we represent the entropic free energy $W_{ent}=4k_BTa\Omega^2/\sqrt{3}$  by an effective magnetostatic energy $W_{mag}=2\pi M^2$ , then we arrive at the entropic charge $q_{ent}=\sqrt{2k_BTa/\pi\sqrt{3}}$  . However, it is clear that the entropic charge derived in this way does not have any local meaning: the separation of two monopoles may either increase or decrease the configuration vector, leading to an attractive or repulsive interaction respectively. It seems that the configuration vector gives a more complete account of entropic forces than does entropic charge, at least in the long wavelength limit.\\
\indent The nonequilibrium thermodynamic theory is generally appropriate to spin ice in the regime of long Debye length at $T < 1 K$. The dynamic correlation function derived here is at least qualitatively similar to that observed in experiment on $\rm Ho_2Ti_2O_7$~\cite{b.17} and we have calculated the long range $B$-field that contributes to the local response sensed by $\mu $SR and NMR. A full description of these experiments would further need to take into account the local magnetic structure. In the future it would be interesting to find other experimental probes that directly isolate the coarse grained fields $H$, $M$ and $B$, to which the present theory should be most directly applicable.

\indent In conclusion, the nonequilibrium thermodynamic approach has the advantages of thermodynamic consistency, physical transparency, and ease of generalization, but the inherent drawbacks of any thermodynamic quasiparticle description: lack of accounting for local details and a range of application that is limited by the accuracy of the quasiparticle picture. We note a recent alternative approach to static correlations ~\cite{b.18} that is useful in connecting with cases where the quasiparticle description breaks down, for example at higher temperature or on diluted spin lattices~\cite{b.19}. Further developments of our approach could include accounting for the chemical kinetic terms in eq.(4), which are relevant to the Wien effect and bound (Bjerrum) pair formation~\cite{b.20}. Also, we have given the results for the case of equal diffusion coefficients, but it is not difficult to generalise to the case where $D_+\neq D_-$ , and in this case there is an additional reason for generation of space magnetic charge: the difference of magnetic monopole conductivities leads to the formation of magnetic space charge arising as a result of their movement. The present approach could also be adapted to describe the coupling of magnetic currents with other thermodynamic forces such as thermal gradients~\cite{b.21} and electric fields~\cite{b.22}, and should be generally useful for predicting consequences of the monopole model that can be tested against experiment.    

\acknowledgements{It is a pleasure to thank Peter Holdsworth for valuable discussions.}



\end{document}